\documentclass[11pt]{article}

\usepackage{amsmath}
\usepackage[dvips]{graphicx}
\usepackage{natbib}

\newenvironment{eqnl}{\begin{equation}}{\end{equation}}
\newenvironment{eqanl}{\begin{eqnarray}}{\end{eqnarray}}

\begin{document}

\title{Matrix Analysis of Tracer Transport}
\author{Peter Mills\\
1159 Meadowlane, Cumberland ON, K4C 1C3, Canada\\
\textit{peteymills@hotmail.com}}

\maketitle

\begin{center}
\includegraphics[width=0.5\textwidth]{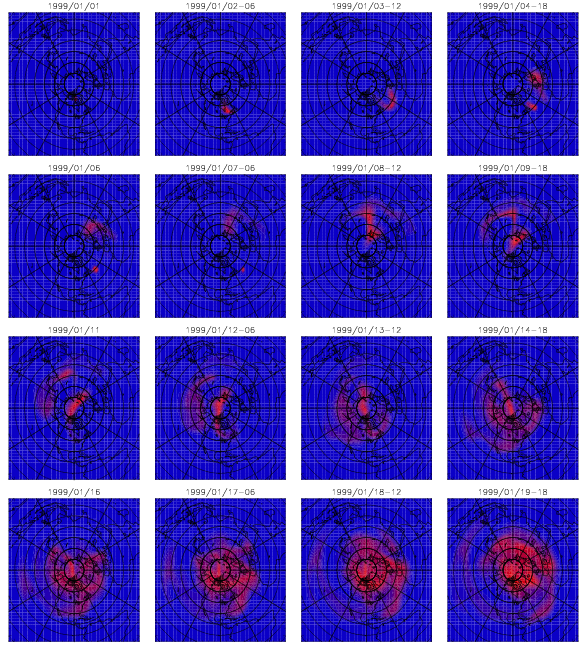}
\end{center}

\pagestyle{myheadings}
\markright{Mills: Tracer Transport}

\begin{abstract}
	We review matrix methods as applied to tracer transport.
Because tracer transport is linear, matrix methods are an ideal
fit for the problem.
A gridded, Eulerian tracer simulation can be approximated
as a system of linear ordinary differential equations (ODEs).
The first-order stretching and deformation of Lagrangian space 
can also be calculated using a system of linear ODEs.
Solutions to these equations are reviewed
as well as special properties.
Using matrices to model tracer transport can also help understand and
improve the stability of numerical solutions.
Detailed derivations are included.

\end{abstract}

\subsection*{Keywords}
\textbf{tracer dynamics, Eulerian transport, numerical analysis, matrix methods, partial differential equations, ordinary differential equations, advection}

\tableofcontents

\section{Introduction}

In \citet{Mills2018}, a method of dynamical tracer interpolation
is introduced
in which the tracer dynamics are represented as a matrix and the largest
principal components correlated with a series of sparse measurements.
The method is called ``principal component proxy tracer analysis''.

Because the processes of tracer advection, as well as the related ones
of local stretching and diffusion, are linear, 
matrix methods represent a powerful and
general set of techniques to apply to these problems.
This review summarizes some of these techniques as well as special properties
of the matrices as used to represent transport processes.  In particular, 
we review methods of solving systems of linear ordinary differential
equations (ODEs) of the form:
\begin{equation}
\frac{\mathrm d \vec r(t)}{\mathrm d t} = A(t) \cdot \vec r(t)
\end{equation}
where $A$ is the instantaneous dynamics, while $\vec r$ is either the tracer,
the stretching of an infinitesimal area in Lagrangian space or the tracer gradient at a single point
in Lagrangian space. 
If $\vec r$ represents the tracer and is a finite vector,
then it is an approximate solution to the following differential equation:
\begin{eqnl}
	\frac{\partial q}{\partial t} = - \vec v \cdot \nabla_{\vec x} q
\end{eqnl}
where $q$ is the {\it tracer mixing ratio}, $\vec x$ is spatial position and the velocity, $\vec v$, is given:
\begin{eqnl}
	\frac{\mathrm d \vec x}{\mathrm t} = \vec v
\end{eqnl}
In this case, $\vec r$ is a discrete approximation of $q$:
\begin{eqnl}
	r_i(t) = q(\vec x_i, t)
\end{eqnl}
where $\lbrace \vec x_i \rbrace$ are the grid points of the model.

The solution can be represented in the form:
\begin{eqnl}
	\vec r=R \cdot \vec r_0
\end{eqnl}
where $\vec r_0=\vec r(t_0)$.
The {\it solution matrix}, $R$, can be broken down in a number of useful ways
and may possess important properties depending upon those of
the {\it evolution matrix}, $A$.

It is hoped that this compilation
can help improve understanding of fluid transport and in particular 
improve both understanding and implementation of principal component
proxy and similar techniques.

\section{Fundamentals}

A trajectory is the motion in time of an infinitesimal packet of fluid
as it is carried along by the flow:
\begin{eqnl}
	\frac{\mathrm d \vec x}{\mathrm d t}=\vec v(\vec x,~t) \label{trajectory_equation}
\end{eqnl}
where $\vec x$ is position, $t$ is time and $\vec v$ is the flow or velocity.
If we integrate this in time, starting at $t_0$ and ending at $t=t_0+\Delta t$, we get a
vector function, call it, $\Phi$:
\begin{equation}
	\vec x=\Phi(\vec x_0,~t_0,~\Delta t)
\label{traj_def}
\end{equation}
where $x_0=x(t_0)$ is the initial position or Lagrangian coordinate.
Unlike in \citet{Ottino1989}, we specify the starting time explicitly instead 
of assuming that $t_0=0$.  This convention will become useful later on, e.g., 
when chaining functions:
\begin{equation}
\vec x=\Phi(\vec x_0,~\Delta t_1+\Delta t_2)=\Phi[\Phi(\vec x_0,~t_0,~\Delta t_1),~t_0 + \Delta t_1,~\Delta t_2]
\label{traj_fun_chaining}
\end{equation}

A flow tracer is a scalar field that follows the flow, typically a dissolved
trace substance, but could also comprise a suspension or in fact any property
of the fluid such as temperature or the velocity field itself.

The change in the total amount of tracer in a fixed volume, $\delta V$, is given
by the flux plus the source term:
\begin{equation}
	\frac{\partial}{\partial t}\int_{\delta V} \rho \mathrm d \vec x=-\oint_A \rho \vec v \cdot \hat n \mathrm d A
	+ \int_{\delta V} \sigma \mathrm d \vec x
	\label{volume_conservation_integral}
\end{equation}
where $\rho$ is the density of the tracer,
$A$ is the area enclosing $\delta V$ and $\sigma$ is the source term, which for the moment,
we will take to be zero but in later analysis will 
be needed both for the diffusion term and for explicit generation and loss terms.

From divergence theorem:
\begin{eqnl}
	\int_{\delta V} \frac{\partial \rho}{\partial t} \mathrm d \vec x=-\int_{\delta V} \nabla \cdot (\rho \vec v)\mathrm d \vec x 
\end{eqnl}
Removing the integrals and expanding the first term on the right side:
\begin{equation}
\frac{\partial \rho}{\partial t} =-\vec v \cdot \nabla \rho - \rho \nabla \cdot \vec v 
\label{mass_conservation_Eulerian}
\end{equation}
The first term is the advection term while the second term is the mass
conservation term and is governed by the 
expansion and contraction of the fluid.  In an incompressible fluid (that is,
$\nabla \cdot \vec v = 0$), this term will be zero.

To go from the {\it Lagrangian} to the {\it Eulerian formulation}, 
we use the following equation for the {\it total derivative}:
\begin{eqnl}
\frac{\mathrm d \rho}{\mathrm d t}
 = \frac{\partial \rho}{\partial t} + \vec v \cdot \nabla \rho
\end{eqnl}
Combining this with Equation (\ref{mass_conservation_Eulerian}):
\begin{equation}
\frac{\mathrm d \rho}{\mathrm d t} =  - \rho \nabla \cdot \vec v
\label{mass_conservation_Lagrangian}
\end{equation}
This is the Lagrangian equation for conservation of mass and governs the total density
of the fluid, that is, fluid density is itself a flow tracer.

Suppose we use {\it mixing ratio} instead of density to track the
tracer:
\begin{eqnl}
	q = \frac {\rho}{\rho_t} \label{mixing_ratio}
\end{eqnl}
where $\rho_t$ is the total density of the fluid.  
Differentiating this with respect to $t$ and 
then substituting Equation (\ref{mass_conservation_Lagrangian}) 
produces the following:
\begin{eqanl}
\frac{\mathrm d q}{\mathrm d t} & = & \frac{1}{\rho_t} \frac{\mathrm d \rho}{\mathrm d t}
	- \frac{\rho}{\rho_t^2}\frac{\mathrm d \rho_t}{\mathrm d t}\\
& = & 0
\end{eqanl}
or, in Eulerian form:
\begin{equation}
\frac{\partial q}{\partial t} = - \vec v \cdot \nabla q
\label{advection_eqn}
\end{equation}
This is the {\it advection equation} \citep{Pedlosky1987}.

\section{Volume deformation}

\label{deformation_section}

Most of this analysis can be found in \citet{Pattanayak2001} and 
\citet{Mills2004}. 
The instantaneous rate of stretching of Lagrangian space is given by the
gradient, or Jacobi matrix, of the velocity, $\nabla \vec v$.
Suppose we perturb a trajectory by a minute amount, $\delta \vec x$.
The Taylor expansion of the time derivative is, to first order:
\begin{equation}
\frac{\mathrm d}{\mathrm d t} (\vec x + \delta \vec x) \approx
\vec v + \nabla \vec v \cdot \delta \vec x \label{taylor_expansion}
\end{equation}
or:
\begin{equation}
\frac{\mathrm d}{\mathrm d t}\delta \vec x \approx \nabla \vec v \cdot \delta \vec x
\label{evolution_error_vector}
\end{equation}
Whether we left multiply or right multiply depends on which convention we adopt
for the application of the gradient or nabla operator, $\nabla$, to a vector,
therefore we write it out component-by-component:
\begin{eqnl}
\nabla \vec v = \left [
\begin{array}{ccc}
\frac{\partial v_x}{\partial x} & \frac{\partial v_x}{\partial y} & \frac{\partial v_x}{\partial z} \\
\frac{\partial v_y}{\partial x} & \frac{\partial v_y}{\partial y} & \frac{\partial v_y}{\partial z} \\
\frac{\partial v_z}{\partial x} & \frac{\partial v_z}{\partial y} & \frac{\partial v_z}{\partial z}
\end{array} \right ]
\end{eqnl}
where $\vec x=[x_1,~x_2,~x_3]=[x,~y,~z]$.

We define $H(\vec x_0,~t_0,~\Delta t)$ as follows:
\begin{equation}
\frac{\mathrm d H}{\mathrm d t} = \nabla \vec v \cdot H
\label{deformation_matrix}
\end{equation}
\begin{eqnl}
H(\vec x_0,~t_0,~0) = I \nonumber
\end{eqnl}
where $I$ is the identity matrix.
This is known as the {\it tangent model} of a dynamical system and applied
to a trajectory is the total stretching of Lagrangian space or 
{\it deformation matrix}.

We can relate $H$ to the integrated trajectory, $\Phi$, as follows:
\begin{equation}
\frac{\mathrm d \Phi}{\mathrm d t} = v(\Phi, ~t)
\end{equation}
\begin{eqnl}
\frac{\mathrm d}{\mathrm d t} 
\nabla_{\vec x_0} \Phi = \nabla \vec v \cdot \nabla_{\vec x_0} \Phi
\label{gradient_trajectory}
\end{eqnl}
or,
\begin{eqnl}
\nabla_{\vec x_0} \Phi = H
\end{eqnl}
Note that:
\begin{equation}
	\delta \vec x = H \cdot \delta \vec x_0
	\label{initial_error_vector}
\end{equation}
where $\delta \vec x_0=\delta \vec x(t=t_0)$.

In conjunction with the equations for the evolution of error vectors, above,
we can derive a set of corresponding equations for the evolution of the tracer
gradient.  Taking the gradient of Equation (\ref{advection_eqn}):
\begin{eqnl}
\frac{\partial}{\partial t} \nabla q = -\nabla q \cdot \nabla \vec v -
		\vec v \cdot \nabla \nabla q
\end{eqnl}
Meanwhile, using the total derivative:
\begin{eqnl}
\frac{\mathrm d}{\mathrm d t} \nabla q = \vec v \cdot \nabla \nabla q \cdot + 
		\frac{\partial}{\partial t} \nabla q
\end{eqnl}
and combining the two:
\begin{equation}
\frac{\mathrm d}{\mathrm d t} \nabla q = -\nabla q \cdot \nabla \vec v
\label{evolution_tracer_gradient}
\end{equation}

In parallel with $H$, we define $H^\prime$:
\begin{equation}
\frac{\mathrm d H^\prime}{\mathrm d t} = -H^\prime \cdot \nabla \vec v
\label{inverse_deformation_matrix}
\end{equation}
\begin{eqnl}
H^\prime(\vec x_0,~t_0,~0) = I
\end{eqnl}
It is easy to show that:
\begin{eqnl}
H^{-1}=H^\prime
\end{eqnl}
by taking the time derivative of $H^\prime \cdot H$:
\begin{eqanl}
\frac{\mathrm d}{\mathrm d t} H^\prime \cdot H & = & 
		H^\prime \cdot \frac{\mathrm d H}{\mathrm d t} +
		\frac{\mathrm d H^\prime}{\mathrm d t} \cdot H\\
		& = & H^\prime \cdot \nabla \vec v \cdot H 
		- H^\prime \cdot \nabla \vec v \cdot H \\
		& = & 0
\end{eqanl}
Again using the properties of derivatives
as in (\ref{initial_error_vector}), we can also show:
\begin{eqnl}
\nabla q = \nabla q_0 \cdot H^\prime
\end{eqnl}

Finally, by defining $\Phi^{-1}(\vec x,~t,~\Delta t)$ as,
\begin{eqnl}
\Phi^{-1}[\Phi(\vec x_0,~t_0,~\Delta t),~t_0,~\Delta t]=\vec x_0
\end{eqnl}
we can show:
\begin{eqnl}
H^\prime=\nabla \Phi^{-1}
\end{eqnl}

\section{Transport map}

\label{map_section}

The dynamics of the tracer, $q$, can be summarized using a linear
{\it transport map}, $Q$:
\begin{equation}
	q(\vec x,~t)=\int_{V} Q(\vec x_0,~\vec x,~t_0,~\Delta t) q(\vec x_0,~t_0) \mathrm d \vec x_0
\label{tracer_map_continuous}
\end{equation}
where $V$ here represents the total volume.
In the absence of diffusion or source terms, this map can be calculated in
at least two ways, by a differential equation: 
\begin{eqnarray}
	\frac{\partial}{\partial t} Q & = & (\vec v \cdot \nabla) Q 
\label{tracer_map_continuous1}\\
Q(\vec x_0,~\vec x,~t_0,~0) & = & \delta(\vec x-\vec x_0) 
\label{tracer_map_continuous2}
\end{eqnarray}
where $\delta$ is the Dirac delta function, and using the trajectory function:
\begin{equation}
Q(\vec x_0,~\vec x,~t_0,~\Delta t) = \delta[\Phi(\vec x_0,~t_0,~\Delta t)-\vec x]
\label{tracer_map_continuous3}
\end{equation}
In the latter case, combination with (\ref{tracer_map_continuous}) produces the
Frobenius-Perron equation. \citep{Ott1993}

The transport map can be approximated by a matrix:
\begin{equation}
\vec q(t) = R \cdot \vec q(t_0)
\label{discrete_tracer_map0}
\end{equation}
where $\vec q$ is a discrete approximation of the tracer configuration:
$q_i = q(\vec x_i)$.  Tracer transport is fully linear and the time evolution
of $R$ is governed by the same mathematics as $H$ and $H^\prime$:
\begin{equation}
\frac{\mathrm d R}{\mathrm d t} = A \cdot R
\label{discrete_tracer_map}
\end{equation}
There are a number of ways to calculate $A$, the most obvious being from an 
Eulerian finite difference scheme, which, in one dimension, would look something
like this:
\begin{equation}
\frac{\mathrm d r_{ij}}{\mathrm d t} = \frac{v(x_j,~t)}{\Delta x_{j-1}+\Delta x_j} (r_{i,j+1} - r_{i,j-1})
\label{simple_finite_difference}
\end{equation}
or,
\begin{eqanl}
a_{i-1,i} & = & - \frac{v(x_i,~t)}{\Delta x_{i-1}+\Delta x_i} \\
a_{i+1,i} & = & \frac{v(x_i,~t)}{\Delta x_{i-1}+\Delta x_i}
\end{eqanl}
Neither representation shows the {\it boundary conditons}.  Typically, $A$
will be either band diagonal or, as in the case of a semi-Lagrangian
scheme, quite close.

We will analysize this system of linear ODEs, as well as the pair of systems
discussed in the previous section, using matrix algebra.  A more thorough
treatment would apply abstract algebra and operator theory to solve the
advection equation, but in many cases the mathematics will be identical.

\section{Matrix solution of systems of linear ODEs} 

\subsection{Analytic solution of the stationary case}

We wish to solve the system of linear ordinary differential equations (ODEs):
\begin{equation}
\frac{\mathrm d \vec r}{\mathrm d t}=A \cdot \vec r
\label{linear_ODE_system_vector_soln}
\end{equation}
where $A$ is the {\it evolution matrix}.
Supposing $A$ has no time dependence, we perform an eigenvector 
decomposion:
\begin{equation}
  A = T \cdot \Lambda \cdot T^{-1}
  \label{eigenvalue_expansion}
\end{equation}
where $T$ is a matrix of right eigenvectors, and $\Lambda$ is a diagonal matrix
of eigenvalues, $\lambda_{ii}=\lambda_i \ge \lambda_{i-1}$.  
The left eigenvectors are contained in $T^{-1}$.
By performing the linear coordinate transformation,
\begin{eqnl}
  \vec r^\prime=T^{-1} \cdot \vec r
\end{eqnl}
hence,
\begin{eqanl}
	\frac{\mathrm d \vec r^\prime}{\mathrm d t} & = &\Lambda \cdot \vec r^\prime \\
	\frac{\mathrm d r_i^\prime}{\mathrm d t} & = &\lambda_i r_i^\prime
\end{eqanl}
the equation is easily solved:
\begin{eqnl}
r^\prime_i = e^{\lambda_i \Delta t} r^\prime_i(t_0)
\end{eqnl}
or, in the un-transformed system:
\begin{eqnarray}
  \vec r & = & \left [T \cdot \exp(\Delta t \Lambda) \cdot T^{-1} \right ] \cdot \vec r_0 
\label{solution_no_time_dependence} \\
& \equiv & \exp(\Delta t A)\cdot \vec r_0
\end{eqnarray}
where $\vec r_0=\vec r(t_0)$ \citep{Robinson2004}.

\subsection{Solving the time-dependent case}

The stationary case is interesting, but what can it tell us about the 
more general case in which $A$ is time-dependent?
We can generalize the problem further as in (\ref{deformation_matrix}) and
(\ref{inverse_deformation_matrix}) so that a matrix, $R$, 
is used in place of the vector, $\vec r$:
\begin{equation}
\frac{\mathrm d R}{\mathrm d t}=A(t) \cdot R
\label{linear_ODE_system_matrix_soln}
\end{equation}
One of the most important properties of the solution is that it
can be decomposed in terms of itself:
\begin{equation}
	R(t_0,~t_n-t_0) = R(t_n,\,\Delta t_n) \cdot R(t_{n-1},\,\Delta t_{n-1}) \cdot R(t_{n-2},\,\Delta t_{n-2}) \cdot \, ...~~ 
	\cdot R(t_0,\,\Delta t_0)
\label{matrix_soln_decomposition}
\end{equation}
where,
\begin{eqnl}
t_n=t_0+\sum_{i=0}^n \Delta t_i
\end{eqnl}
and we have used the convention, begun
in Equation (\ref{traj_def}) of making $R$ a function both of the
initial time and of the subsequent time interval.  
It follows that $R(t, 0)=I$.

The vector solution is given by a product of the initial vector with
the matrix solution:
\begin{eqnl}
	\vec r(t_0+\Delta t)=R(t_0, \Delta t) \cdot \vec r(t_0)
\end{eqnl}
which can be show by substitution back into (\ref{linear_ODE_system_vector_soln}).

Each element in the decomposition in (\ref{matrix_soln_decomposition})
may be approximated by the stationary solution in the limit as the time step
approaches zero:
\begin{equation}
	\lim_{\Delta t \rightarrow 0} \left \lbrace
R(t, ~ \Delta t) = \exp \left [ A(t) \Delta t \right ]
\right \rbrace
\end{equation}

\subsection{Negative and left-multiplied cases}

Consider the case in which the order of the factors on the right-hand-side (RHS) of Equation 
(\ref{linear_ODE_system_vector_soln}) are reversed (left-multiply vs. right-multiply):
\begin{eqnl}
\frac{\mathrm d \vec r}{\mathrm d t} = \vec r \cdot A
\end{eqnl}
This case is particularly important, since Equation (\ref{evolution_error_vector})
when rearranged in this way becomes the vorticity equation \citep{Acheson1990}.

We start with the analytic solution of the stationary case:
\begin{eqanl}
  A & = & T \cdot \Lambda \cdot T^{-1} \\
  \frac{\mathrm d \vec r}{\mathrm d t} & = & \vec r \cdot T \cdot \Lambda \cdot T^{-1} \\
  \frac{\mathrm d}{\mathrm d t} (\vec r \cdot T) & = & (\vec r \cdot T) \cdot \Lambda \\
  \vec r \cdot T & = & \vec r_0 \cdot T \cdot \exp (\Lambda t) \\
	\vec r & = & \vec r_0 \cdot T \cdot \exp (\Lambda t) \cdot T^{-1}
\end{eqanl}
In other words, the solution is the same, but the initial conditions are
left-multiplied instead of right multiplied, or equivalently, the whole thing 
could be transposed.

In the time-dependent case each element of the solution is in reverse order,
not just the initial conditions:
\begin{eqnl}
	\vec r(t_n) = \vec r_0 \cdot R^*(t_0,\Delta t_0) \cdot R^*(t_1, \Delta t_1) \cdot \cdot R^*(t_2, \Delta t_2) ~ ... 
~ \cdot R^*(t_n,\Delta t_n)
\end{eqnl}
where $R^*$ is a solution to the equation:
\begin{eqanl}
\frac{\mathrm d}{\mathrm d t}R^*(t_0,~t) & = & R^*(t_0, ~t) \cdot A(t) \\
R^*(t_0, ~ 0) & = & I
\end{eqanl}
and,
\begin{eqnl}
R^*(t,~\Delta t \rightarrow 0) = R(t,~ \Delta t ) 
 = \exp \left [ A(t) \Delta t \right ]
\end{eqnl}

For the case of a negative RHS:
\begin{eqnl}
\frac{\mathrm d \vec r}{\mathrm d t} = - A \cdot \vec r
\end{eqnl}
we can show that the stationary solution is the inverse of that for a
positive RHS:
\begin{eqanl}
  \vec r & = & T \cdot \exp (-\Lambda t) \cdot T^{-1} \cdot \vec r \\
	& = & T \cdot \left [ \exp (\Lambda t) \right ]^{-1} \cdot T^{-1} \cdot \vec r \\
 & = & \left [T \cdot \exp (\Lambda t) \cdot T^{-1} \right]^{-1} \cdot \vec r 
\end{eqanl}
while for the time-dependent case, we have:
\begin{eqnl}
	\vec r \approx R^{-1}(t_n,\Delta t_n) \cdot R^{-1}(t_{n-1},\Delta t_{n-1}) \cdot R^{-1}(t_{n-2}, \Delta t_{n-2}) \cdot ~ ... 
~ \cdot R^{-1}(t_0,\Delta t_0) \cdot \vec r_0
\end{eqnl}
assuming that each $\Delta t_i$ is small enough that $R$ approximates the 
stationary case.
This provides an alternative derivation for the solution of the negative,
transposed (left-multiplied) case which we saw already in (\ref{evolution_tracer_gradient}):
\begin{eqnl}
\frac{\mathrm d R}{\mathrm d t} = -\vec r \cdot A
\end{eqnl}
which is given by:
\begin{eqnarray}
\vec r(t_n) & \approx & \vec r_0 \cdot R^{-1}(t_0,\Delta t_0) \cdot R^{-1}(t_1, \Delta t_1) \cdot ~ ... 
~ \nonumber \\
& & ~~~~~~~...~\cdot R^{-1}(t_{n-1},\Delta t_{n-1}) \cdot R^{-1}(t_n,\Delta t_n) \\
& = & \vec r_0 \cdot R^{-1}(t_0, t_n-t_0)
\end{eqnarray}
Where $R$ is the solution to Equation (\ref{linear_ODE_system_matrix_soln}).

\section{SVD and the Lyapunov spectrum}

The singular value decomposition of a matrix is given as:
\begin{equation}
R=U\cdot S\cdot V^T
\label{SVD_def}
\end{equation}
where $R$ is an $[m \times n]$ matrix, $U$ is an $[m \times n]$ orthogonal
matrix, $S$ is an $[n \times n]$ diagonal matrix of {\it singular values}
($s_{ii}=s_i\ge s_{i+1}$) and $V$ is an $[n \times n]$ orthogonal matrix
\citep{Press_etal1992}.

$U$ and $V^T$ are also termed the left and right {\it singular vectors}, 
respectively and are normally calculated through eigenvalue analysis:
\begin{eqnarray}
	R\cdot R^T \cdot U & = & U \cdot S^2 \label{left_SV_eigenproblem}\\
	R^T \cdot R \cdot V & = & V \cdot S^2 \label{right_SV_eigenproblem}
\end{eqnarray}
Typically, only one of $U$ or $V$ is calculated and then the other by projection
onto $R$.  Which one is calculated first is best determined by whether $m$
is greater than or less than $n$.  Equation (\ref{SVD_def}) assumes that 
$m>n$. 
For all the problems discussed in this review, $m=n$.

Because both $R^T \cdot R$ and $R \cdot R^T$ are symmetric,
the singular values,
$\lbrace s_i \rbrace$, are always real.
Moreover, the eigenvectors in $U$ and $V$ form an orthogonal set spanning the
space and are normalized so that,
$U^T \cdot U = V^T \cdot V = I$.
Matrices with this property are often called {\it ortho-normal}.

Assuming that $R$ is an integrated tangent model as in 
(\ref{deformation_matrix}), (\ref{inverse_deformation_matrix})
and (\ref{discrete_tracer_map}), 
the Lyapunov exponents are defined as the time averages of the logarithms
of the singular values in the limit as time goes to infinity \citep{Ott1993}:
\begin{equation}
h_i=\lim_{\Delta t \rightarrow \infty} \frac{1}{\Delta t} \log s_i
\label{Lyap_def}
\end{equation}

If there is any significant difference between the largest and next largest
Lyapunov exponents, the largest singular value will come to dominate the matrix
as it evolves forward in time.  Thus:
\begin{eqnl}
r_i(t \rightarrow \infty) = u_{i1} s_1 \sum_j v_{j1} r_j(t_0)
\end{eqnl}
and:
\begin{equation}
|\vec r(t \rightarrow \infty)| = |\vec r_0| e^{h_1 \Delta t}
\label{large_Lyap}
\end{equation}
The Lyapunov exponent is often somewhat incorrectly defined as
(\ref{large_Lyap}) above \citep{Ott1993}.

\section{Special properties}

In many cases of the problems discussed in Section \ref{deformation_section}
and Section \ref{map_section}, the evolution matrix, $A$, will have
special properties that will affect the solution.
Since each element in (\ref{matrix_soln_decomposition}) approaches the
non-time-dependent solution in the limit as $\Delta t_i\rightarrow 0$, 
we can sometimes use the properties of the non-time-dependent solution to
reason about those of the time-dependent one.

\subsection{Volume conservation}

In the solution of (\ref{deformation_matrix}) the evolution matrix
is equal to the gradient of the velocity,
$A=\nabla \vec v$, while the velocity field, $\vec v$, is frequently non-divergent,
$\nabla \cdot \vec v=0$ or $\mathrm{Tr}(A)=0$.
It can be shown that 
if the trace of a matrix is zero, then the eigenvalues sum to zero
\citep{Anton1987}.
Here we outline the proof.

Begin by writing the characteristic polynomial as a power series:
\begin{eqnl}
	\lambda^n + k_1 \lambda^{n-1} + k_2 \lambda^{n-2} + k_3 \lambda^{n-3} + ~ ... ~ = 0
\end{eqnl}
The first coefficient, $k_1$, is given:
\begin{eqnl}
	k_1 = a_{11} + a_{22} + a_{33} + ~ ... ~ = \mathrm{Tr}(A)
\end{eqnl}
We can also write the characteristic polynomial
in terms of the roots:
\begin{eqanl}
	& (\lambda - \lambda_1)(\lambda - \lambda_2)(\lambda-\lambda_3) ~ ... \\
	= & \lambda^n + (\lambda_1 + \lambda_2 + \lambda_3 + ~ ...) \lambda^{n-1} + ...
\end{eqanl}
Thus:
\begin{eqnl}
	k_1 = \mathrm{Tr}(A) = \sum_i \lambda_i = 0
\end{eqnl}

The determinant of the stationary solution matrix,
$R=T\cdot\exp(\Delta t\Lambda)\cdot T^{-1}$, in (\ref{solution_no_time_dependence}) is:
\begin{eqanl}
	|R| & = & |T||\exp(\Delta t\Lambda)||T^{-1}| \\
	    & = & |T| \left [ \prod_i \exp(\Delta t \lambda_i) \right ] \frac{1}{|T|} \\
& = & \exp\left(\Delta t \sum_i \lambda_i\right) \\
& = & 1
\end{eqanl}
In other words, the solution, in this case, is {\it volume-conserving}:
volumes in the space, e.g. as calculated by the determinant of a set of 
solution vectors
spanning the space, are conserved.  
This is known as Liouville Theorem \citep{Thornton2003}.

The result generalizes to the time-dependent case since the solution can
always be decomposed as an infinite product of infinitessimally small 
integrations of the evolution matrix each with eigenvalue zero.
Like the eigenvalues, the Lyapunov exponents will also sum to zero:
\begin{eqanl}
|R| =	|U||S||V^T| & = & 1 \\
	\prod_i s_i & = & 1 \\
	\prod_i \exp(\Delta t h_i) & = & 1 \\
	\sum_i h_i & = & 0
\end{eqanl}

\subsection{Mass and length conservation}

Suppose $R$ has the property that it preserves lengths when applied to
a vector:
\begin{equation}
|R\cdot \vec q| = |\vec q|
\label{length_preservation}
\end{equation}
thus the rate of change of the vector will always be perpendicular to it:
\begin{eqanl}
\frac{\mathrm d}{\mathrm d t}|\vec q| & = & 
	\nabla |\vec q| \cdot \frac{\mathrm d \vec q}{\mathrm d t} \\
	&=& \frac{\vec q}{|\vec q|} \cdot A \cdot \vec q \\
	&=& 0
\end{eqanl}
\begin{equation}
	\vec q \cdot A \cdot \vec q = 0 
	\label{length_conserving_A}
\end{equation}
$A$ transforms a vector so that the result is always orthogonal to the untransformed value.
By separating the diagonal and off-diagonal components in (\ref{length_conserving_A},
\begin{eqnl}
\sum_i \sum_j a_{ij} q_i q_j = \sum_{i=1}^n a_{ii}q_i^2 + \sum_{i=1}^n \sum_{j=i+1}^n (a_{ij} q_i q_j + a_{ji} q_i q_j) = 0
\end{eqnl}
we can show that it has the following
properties:
\begin{eqanl}
a_{ii} & = & 0 \\
a_{ij}+a_{ji} & = & 0
\end{eqanl}

Meanwhile, $R$ is a rotation.  All the singular values of $R$ will be $1$:
\begin{equation}
	\vec q \cdot R^T \cdot R \cdot \vec q = \vec q \cdot \vec q \label{length_conservation}
\end{equation}
This implies:
\begin{eqnarray}
R^T \cdot R \cdot \vec v & = & s^2 \vec v \nonumber \\
	s & = & 1 \label{ortho_eigen}
\end{eqnarray}
or, more simply:
\begin{eqnl}
	R^T \cdot R \cdot \vec v = \vec v
\end{eqnl}
Because it is a rotation, it will be ortho-normal with a determinant of 1.

The tracer mapping, $Q$, as defined in 
(\ref{tracer_map_continuous1}) and (\ref{tracer_map_continuous2}),
that is, without diffusion, will fulfill these properties.  Discrete mappings
can only approximate them and by necessity always include some diffusion.

A more important property of tracer advection is that the total substance
remains constant, thus:
\begin{equation}
\sum_i q_i = const.
\label{mass_conservation}
\end{equation}
This will be true only if $q$ is a volume-mixing-ratio
and
the flow is non-divergent or
$q$ is a density and the total volume of the fluid neither expands
nor contracts, otherwise the formula will be more complicated.  
It also assumes an equal-volume grid.
Many, if not most, flows in real fluids are approximately 
non-divergent, especially when considered over long time scales.

If the total substance is constant, its rate of change will be zero:
\begin{eqnl}
\frac{\mathrm d}{\mathrm d t}\sum_i q_i = 0
\end{eqnl}
Continuing,
\begin{eqanl}
\sum_i \sum_j r_{ij} q_j & = & \sum_j q_j \\
\sum_j q_j \left ( \sum_i r_{ij} - 1 \right ) & = & 0
\end{eqanl}
and:
\begin{eqanl}
\sum_i \frac{\mathrm d q_i}{\mathrm d t} & = & 0 \\
\sum_i \sum_j a_{ij} q_j & = & 0 \\
\sum_j q_j \sum_i a_{ij} & = & 0 
\end{eqanl}
Therefore:
\begin{eqnarray}
\sum_i r_{ij} & = & 1 
\label{column_sums_to_one}\\
\sum_i a_{ij} & = & 0
\label{column_sums_to_zero}
\end{eqnarray}

For grids that aren't equal-volume, we apply the following transformations:
\begin{eqnl}
  \tilde q_i = w_i q_i
\end{eqnl}
where $w_i$ is a weight accounting for the relative differences 
in volume at the $i$th grid point.
The relevant matrices will also need to be transformed:
\begin{eqanl}
  \tilde r_{ij} & = & \frac{w_i}{w_j} r_{ij} \\
  \tilde a_{ij} & = & \frac{w_i}{w_j} a_{ij}
\end{eqanl}
The same constraints in 
(\ref{column_sums_to_one}) and (\ref{column_sums_to_zero}) 
now apply to the transformed matrices, $\tilde R=\lbrace \tilde r_{ij} \rbrace$
and $\tilde A=\lbrace \tilde a_{ij} \rbrace$.
If the weights are given time dependence, a similar method can be
applied to account for local changes in density in non-divergent flows.

As pointed out already, a discrete tracer mapping will always require some 
amount of diffusion.  This means that the tracer configuration will 
tend towards a uniform distribution over time, 
that is, it will ``flatten out''.  We can
show that, given the the constraint in (\ref{mass_conservation}), 
a tracer field with all the same values
 has the smallest magnitude.  Suppose there are only two elements in the 
tracer vector, $\vec q=\lbrace q,~q \rbrace$.  The magnitude of the vector is:
\begin{eqnl}
|\vec q|=\sqrt{q^2+q^2}=\sqrt{2} q
\end{eqnl}
Now we introduce a separation between the elements, $2\Delta q$, that 
nonetheless keeps the sum of the elements constant:
\begin{eqanl}
|q+\Delta q,~q-\Delta q| & = & \sqrt{(q+\Delta q)^2+(q-\Delta q)^2} \\
& = & \sqrt{2}\sqrt{q^2+(\Delta q)^2} \ge \sqrt{2} q
\end{eqanl}
This will generalize to higher-dimensional vectors.  In general, we can
say that:
\begin{equation}
\vec q \cdot R^T \cdot R \cdot \vec q \le \vec q \cdot \vec q
\label{tracer_map_inequality}
\end{equation}
Implying that for the eigenvalue problem,
\begin{eqnarray}
R^T \cdot R \cdot \vec v & = & s^2 \vec v \nonumber\\
s^2 & \le & 1 \label{SV_inequality}
\end{eqnarray}
The full proof is outlined below.
This further shows that the Lyapunov exponents are all
either zero or negative with the largest equal to 0.  This has been shown
numerically in \citet{Mills2018}.

To prove (\ref{SV_inequality}) from (\ref{tracer_map_inequality}), we first
expand $\vec q$ in terms of the right singular values, 
$\lbrace \vec v_i \rbrace$:
\begin{eqnl}
	\vec q = \sum_i c_i \vec v_i
\end{eqnl}
where $\lbrace c_i \rbrace$ are a set of coefficients.
Substituting this into the left-hand-side of (\ref{tracer_map_inequality}):
\begin{eqanl}
	\vec q \cdot R^T \cdot R \cdot \vec q & = & \left ( \sum_i c_i \vec v_i \right ) \cdot \left (\sum_i c_i s_i^2 \vec v_i \right ) \\
   & = & \sum_i \sum_j c_i c_j s_i^2 \vec v_i \cdot \vec v_j \\
   & = & \sum_i \sum_j c_i c_j s_i^2 \delta_{ij} \\
	  & = & \sum_i c_i^2 s_i^2
\end{eqanl}
where $\delta$ is the Kronecker delta.
Similarly, we can show that:
\begin{eqnl}
	\vec q \cdot \vec q = \sum_i c_i^2
\end{eqnl}
If we assume that $s_i \le 1$ for every $i$, then:
\begin{equation}
	\sum_i c_i^2 s_i^2 \le \sum_i c_i^2 
	\label{diffusive_inequality_in_terms_of_SVs}
\end{equation}
since each term on the left-hand-side is less-than-or-equal-to the
corresponding term on the right-hand-side. 
Note that in order for the inequality in 
(\ref{diffusive_inequality_in_terms_of_SVs}) to be broken, at least one
singular value must be greater-than one.
Therefore (\ref{tracer_map_inequality}) is true for every $\vec q$
if-and-only-if (\ref{SV_inequality}) is true for every $s$.
In the language of set theory and first-order logic:
\begin{eqnl}
	\forall \vec q \in \Re^n ~ (\vec q \cdot R^T \cdot R \cdot \vec q \le \vec q \cdot \vec q) \iff \forall s \in \Re | ~R^T \cdot R \cdot \vec v = s^2 \vec v ~ (s \le 1)
\end{eqnl}
A similar argument will prove (\ref{length_conservation})
iff (\ref{ortho_eigen}).


\ifdefined\interesting
\section{Volume deformation versus the transport map}

Since the deformation matrices,
$H$ and $H^\prime$, and the transport map, $Q$, are all derived from the
transport equations, and their time evolution has the same mathematical form,
we would expect them to be closely related. 
The $H$ matrices describe the local deformation of Lagrangian space 
while $Q$ describes the global deformation.
Since $\delta \left [\Phi(\vec x_0) - \vec x \right ]=\delta \left [\vec x_0 - \Phi^{-1}(\vec x) \right ]$,
we can derive (\ref{inverse_deformation_matrix}) from
(\ref{tracer_map_continuous}) and (\ref{tracer_map_continuous3}) by taking 
the gradient of $q$:
\begin{eqanl}
	\nabla_{\vec x} q & = & - \int_V \vec \delta^\prime \left [\vec x_0 - \Phi^{-1}(\vec x) \right ] q_0 (\vec x_0) 
	\mathrm d \vec x_0 \cdot \nabla_{\vec x} \Phi^{-1} \\
	& = & - \nabla_{\vec x_0} q_0 \cdot H^\prime
\end{eqanl}
where $\vec \delta^\prime$ is the generalized vector derivative of the 
Dirac delta function.

Once we admit diffusion, the final tracer gradient is no longer locally 
dependent on the gradient at the starting point, but rather depends on the
whole initial tracer configuration:
\begin{eqnl}
	\nabla_{\vec x} q = \int_V \nabla_{\vec x} Q(\vec x_0, \vec x) q_0 (\vec x_0) \mathrm d \vec x_0
\end{eqnl}

As with $\Phi$ and $\Phi^{-1}$, 
there will exist an inverse transport map, $Q^{-1}$, from which we can derive
$H$. In the absence of diffusion, the mathematics will be identical whether
we move forward or backward in time.
That is, for every $q_0$:
\begin{eqnl}
	\int_V Q^{-1}(\vec x, \vec x_0, t_0+\Delta t, \Delta t) 
	\int_V Q(\vec x_0^\prime, \vec x, t_0, \Delta t) 
	q_0(\vec x_0^\prime, t_0) 
	\mathrm d \vec x_0^\prime \mathrm d \vec x = q_0(\vec x_0, t_0)
\end{eqnl}
or:
\begin{eqanl}
	\int_V  Q^{-1}(\vec x, \vec x_1, t_0+\Delta t, \Delta t) Q(\vec x_2, \vec x, t_0+\Delta t, \Delta t) \mathrm d \vec x = \delta(\vec x_1 - \vec x_2)
\end{eqanl}
For the non-diffusive case, $Q$ is a rotation and therefore orthogonal.
We construct the inverse by simply reversing the roles of the variables:
$Q^{-1}(\vec x, \vec x_0) = Q(\vec x_0, \vec x)$, 
the equivalent of the matrix transpose in the discrete case.

Of particular interest in the diffusive case
is what happens to these maps in the limit as time
goes to infinity, $\Delta t \rightarrow \infty$.
By SVD:
\begin{eqnl}
	\lim_{\Delta t \rightarrow \infty} H^\prime = 
	s_{h^\prime1} \vec h_f^\prime \otimes \vec h_0^\prime
\end{eqnl}
where $\otimes$ is the outer product and $s_{h^\prime1}$ is the largest 
singular value. The tracer gradient becomes:
\begin{eqnl}
	\nabla_{\vec x} q = s_{h^\prime1} \vec h_f^\prime
	\nabla_{\vec x_0} q_0 \cdot \vec h_0^\prime
\end{eqnl}

Once we generalize SVD to continuous distributions, a similar thing will
be seen to happen with the transport map:
\begin{equation}
	\lim_{\Delta t \rightarrow \infty} Q(\vec x_0, \vec x) =
	Q_f (\vec x) Q_0 (\vec x_0)
	\label{transport_map_separable}
\end{equation}
In other words, it has become separable and:
\begin{eqnl}
	\nabla_{\vec x} q = \nabla_{\vec x} Q_f \int_V Q_0 (\vec x_0) q_0 (\vec x_0) \mathrm d \vec x_0
\end{eqnl}
Note that no singular value has been included since we saw from
(\ref{SV_inequality}) that it should be close to unity.
(This was demonstrated numerically in \citet{Mills2018}.)


\fi

\section{Practical considerations}

\label{practical_considerations}

The traditional method of testing Eulerian tracer simulations for stability
is von Neumann analysis \citep{Anderson1994}.
Equation (\ref{simple_finite_difference}) provides a simple, finite
difference method for
calculating the evolution matrix, $A$, for an Eulerian tracer simulation.
Assuming we've obeyed the Courant-Friedrichs-Lewy (CFL) criterion 
\citep{Courant_etal1967} for choosing
the time-step (which itself is derived from von Neumann analysis), von Neumann
analysis shows this method to be unconditionally stable.

In fact the method is quite naive and if applied to any real fluid flow would 
quickly overflow.
This is because shear flows and mixing cause the tracer gradient to grow without bound.
To fix this, some amount of diffusion must be added.  Here is the 
{\it advection-diffusion} equation:
\begin{equation}
\frac{\partial q}{\partial t} = - \vec v \cdot \nabla q + \nabla \cdot D \cdot \nabla q
\label{advection_diffusion}
\end{equation}
where $D$ is the diffusivity tensor.
Here is its translation, in one dimension, to a second-order, centred,
finite-difference equation with uniform spatial grids:
\begin{eqnarray}
\frac{\partial q_i}{\partial t} & = & \frac{v(q_{i+1} - q_{i-1})}{2 \Delta x} +
	\frac{d (q_{i-1} + q_{i+1} - 2 q_i)}{\Delta x^2} \\
& = & \left (- \frac{v}{2 \Delta x} + \frac{d}{\Delta x^2} \right ) q_{i-1} -
	\frac{2 d}{\Delta x^2} q_i + 
	\left (\frac{v}{2 \Delta x} + \frac{d}{\Delta x^2} \right ) q_{i+1} \label{finite_difference_diffusion}
\end{eqnarray}
where $d$ is a scalar diffusion coefficient.
Expressed as elements of a matrix:
\begin{eqanl}
a_{i,i-1} & = & \left (- \frac{v}{2 \Delta x} + \frac{d}{\Delta x^2} \right ) \\
	a_{i,i} & = & -\frac{2 d}{\Delta x^2} \\
a_{i,i+1} & = & \left (\frac{v}{2 \Delta x} + \frac{d}{\Delta x^2} \right )
\end{eqanl}
In order to prevent the simulation from overflowing, $d$ will have to be tuned.
Generalization to two or more dimensions is straightforward.
Since the simulation is discrete, the actual order of points in the
vector or matrix is arbitrary.

Semi-Lagrangian methods perform a back-trajectory from each of the Eulerian
points in the simulation, then interpolate the new value.  Rather than
directly integrating the tracer values, 
the interpolation coefficients are stored immediately
in the transport matrix, $R(t_i,~\Delta t_i)$, skipping the integration
step.
Each Eulerian time-step produces one element in the decomposition of the
larger transport matrix $R(t_0,~t_n-t_0)$.

Semi-Lagrangian methods are unconditionally stable and not subject to the
CFL criterion.
Both the final transport map, and each member of its decomposition normally have
the following properties:
\begin{itemize}
\item every element is between 0 and 1:
\begin{equation}
0 \le r_{ij} \le 1
\label{element_on_unit_interval}
\end{equation}
\item rows sum to 1:
\begin{equation}
\sum_j r_{ij} = 1
\label{rowssumtoone}
\end{equation}
\end{itemize}
The first condition holds for linear interpolation methods which will have
a maximum of $2^N$ non-zero elements per row, 
where $N$ is the number of dimensions. 
Some kernel-smoothing methods include negative coefficients.

In addition, we can approximate the evolution matrix to second order:
\begin{eqnl}
A(t_i+\Delta t/2) \approx \frac{1}{\Delta t} \left [R(t_i,~\Delta t) - I \right ]
\end{eqnl}

\subsection{Criteria for stability}

If the transport map, $R$, satisfies both Equation (\ref{column_sums_to_one})
and the inequality in (\ref{element_on_unit_interval}) then it is mathematically
equivalent to a conditional probability.
In addition, if we normalize the tracer such that $\sum_i q_i=1$
then it too is equivalent to a probability.

By translating Equation (\ref{length_preservation}) into an inequality, 
we arrive at a pair of criteria for the evolution matrix:
\begin{eqanl}
a_{ii} & \le & 0 \\
a_{ij}+a_{ji} & \le & 0
\end{eqanl}
in agreement with our definition of a strictly diffusive transport map in
(\ref{tracer_map_inequality}).

\section{Adding sources and sinks}

We wish to add a source term, $\vec \sigma$, to the matrix tracer model
described in the previous sections:
\begin{eqnl}
\frac{\mathrm d \vec q}{\mathrm d t} = A(t) \cdot \vec q + \vec \sigma(t)
\end{eqnl}
Integrating to a discrete, first order approximation:
\begin{eqanl}
  \vec q(t_n) 
  & \approx & \Delta t_{n-1} \sigma(t_{n-1}) + \nonumber \\
  & & R(t_{n-1}, ~ \Delta t_{n-1}) \cdot [(\Delta t_{n-2} \vec \sigma(t_{n-2}) + \nonumber \\
  & & R(t_{n-2}, ~ \Delta t_{n-2}) \cdot [\Delta t_{n-3} \vec \sigma(t_{n-3}) + \nonumber \\
  & & ~... + \nonumber \\
  & & R(t_2, ~ \Delta t_2) \cdot [\Delta t_1 \vec \sigma(t_1) + \nonumber \\
  & & R(t_1, ~ \Delta t_1) \cdot [\Delta t_0 \vec \sigma(t_0) + \nonumber \\
  & & R(t_0, ~ \Delta t_0) \cdot q(t_0) ]]...]]
\end{eqanl}
and multiplying through:
\begin{eqanl}
\vec q(t_n) 
  & \approx & R(t_0, ~ t_n-t_0) \cdot \vec q(t_0) + \nonumber \\
  & & \Delta t_0 R(t_1, ~ t_n-t_1) \cdot \vec \sigma(t_0) + \nonumber \\
  & & \Delta t_1 R(t_2, ~ t_n-t_2) \cdot \sigma(t_1) + \nonumber \\
  & & ~...~+ \nonumber \\ 
  & & \Delta t_{n-3} R(t_{n-2}, ~ t_n-t_{n-2}) \cdot \vec\sigma(t_{n-3}) + \nonumber \\
  & & \Delta t_{n-2} R(t_{n-1}, ~ \Delta t_{n-1}) \cdot \vec \sigma(t_{n-2}) + \nonumber \\
  & & \Delta t_{n-1} \sigma(t_{n-1}) \\
  & = & R(t_0,~t_n-t_0) \cdot \vec q(t_0) + 
  \sum_{i=1}^{n} \Delta t_{i-1} R(t_i,~t_n-t_i) \cdot \vec \sigma(t_{i-1})
\end{eqanl}

\subsection{Diffusion}

Unlike external sources and sinks, we can add diffusion to the matrix model 
without fundamentally changing it or its linearity. 
Consider the diffusion equation in one dimension:
\begin{equation}
  \frac{\partial q}{\partial t}=d \frac{\partial^2 q}{\partial x^2}
  \label{heat_equation}
\end{equation}
where $d$ is the diffusion coefficient.
Solving by separation of variables:
\begin{equation}
  q=q_0 e^{-d \omega^2 t} e^{i \omega x}
  \label{diffusion_solution}
\end{equation}
where $\omega$ is the angular frequency of a stationary wave
and $q_0$ is its initial amplitude.
In other words, if we decompose a function using Fourier analysis, the speed
with which each component will decay is proportional to the square of the
angular frequency times the diffusion coefficient \citep{Cannon1984}.

Now consider an initial point distribution centred at the origin 
which will be evenly spread
in frequency space, leaving only the first factor in 
Equation (\ref{diffusion_solution}). 
This is a Gaussian function and transformed
back into regular ($x$) space returns another Gaussian:
\begin{equation}
	\frac{1}{2 \pi} \int e^{-d \omega^2 t} e^{i \omega x} \mathrm d \omega = \frac{e^{-\frac{x^2}{4 d t}}}{2 \sqrt{\pi d t}}
	\label{Gaussian_diffusion}
\end{equation}
Thus, by convolution theorem \citep{Katznelson1976}, 
applying diffusion to a function for time $t$ 
with diffusion coefficient $d$ 
is equivalent to convolution with a Gaussian of width $\sqrt{2 d t}$.

A convenient property of Gaussian functions 
is that they are dimensionally separable
making results easy to generalize from one to multiple dimensions:
\begin{eqanl}
	\left (2 \pi \right )^{-N/2} e^{-| \vec x |^2/2} 
	& = & \left ( 2 \pi \right )^{-N/2} \exp \left ( \frac{1}{2} \sum_{i=1}^N x_i^2 \right ) \\
	& = & \prod_{i=1}^N \frac{e^{x_i^2/2}}{\sqrt{2 \pi}}
\end{eqanl}
where $N$ is the number of dimensions.
The solution to Equation (\ref{heat_equation}) in multiple dimensions is likewise
separable:
\begin{eqnl}
	q=q_0 \prod_{j=1}^N e^{-d \omega_j^2 t} e^{i \omega_j x_i} = q_0 e^{-d|\vec \omega|^2 t} e^{i \vec \omega \cdot \vec x}
\end{eqnl}
where $\vec \omega=\lbrace \omega_j \rbrace$ is the vector angular frequency
of the stationary wave.

There are a number of different approaches to adding diffusion.
If the transport matrix, $R$, is integrated from an ``evolution matrix'',
$A$, as in Equation (\ref{linear_ODE_system_matrix_soln}), then a diffusion
term can be included in it directly, 
as in the finite difference example in (\ref{finite_difference_diffusion}).

In a semi-Lagrangian scheme, the transport matrix is generated directly
by gathering the interpolation coefficients, so diffusion must be added
separately.
Since interpolation of back-trajectories provides a small amount of implicit
diffusion, we should first calculate how much that is before adding it in
more explicitly.
For a linearly-interpolated semi-Lagrangian scheme in one dimension, 
if we assume the back-trajectory falls directly in
the centre of the grid, then the coefficients form a truncated kernel each
at half-maximum of the Gaussian in (\ref{Gaussian_diffusion}):
\begin{equation}
e^{-\frac{\left(\Delta x/2 \right)^2}{4d_0 \Delta t}} = 1/2
\end{equation}
where $\Delta x$ is the grid spacing and $\Delta t$ is the Eulerian time step,
and $d_0$ is the ``implicit'' diffusion.
Since the series is truncated and the normalization is the sum of the 
interpolation coefficients, the normalization coefficient in the denominator
of (\ref{Gaussian_diffusion}) has been left off. 
Solving for the approximate diffusion coefficient:
\begin{eqnarray}
	d_0 & = & \frac{\Delta x^2}{16 \ln 2 \Delta t} \\
	& = & \frac{\Delta x^2}{11.09 \Delta t}
\end{eqnarray}
Because of the dimensional separability of the normal distribution,
and because of how the number of points used per interpolate scales, 
the result generalizes to higher dimensions.
If the desired diffusion is not too large, the diffusion coefficient can
be set by varying the time step.

The convolution of a Gaussian with another Gaussian is a third Gaussian whose variance is the sum of the variances of the other two:
\begin{eqnl}
	\frac{1}{2 \pi \sigma_1 \sigma_2} \int e^{-\frac{{x^\prime}^2}{2 \sigma_1^2}} e ^{-\frac{(x^\prime - x)^2}{2 \sigma_2^2}} \mathrm d x^\prime
	= \frac{e^{-\frac{x^2}{2(\sigma_1^2 + \sigma_2^2)}}}{\sqrt{2 \pi (\sigma_1^2 + \sigma_2^2)}}
\end{eqnl}
Thus, if we wish to add diffusion through a smoothing matrix,
applied after each time step, we calculate the diffusion coefficient, $d_s$,
to use for the Gaussian kernel from the desired diffusion coefficient, $d$,
less the implicit diffusion coefficient, $d_0$:
\begin{eqnl}
	d_s = d - \frac{\Delta x^2}{11.09 \Delta t}
\end{eqnl}
The advantage to this scheme is that varying amounts of diffusion can be
added as needed after integrating the transport matrix.
Alternatively, rather than linear interpolation, the interpolation
coefficients can be calculated in the same way as each row of the smoothing
matrix: based on a Gaussian kernel according to (\ref{Gaussian_diffusion}).

\section{Conclusions}

Eulerian tracer dynamics can be expressed as the product between a matrix,
which captures the tracer dynamics, and a vector, which represents the tracer
configuration.  If the matrix is considered as a multi-dimensional function
taking as parameters the start time 
in addition to the integration time, then it may be decomposed in terms of
itself by splitting the integration into shorter intervals.  
By considering the tracer in this manner we can take advantage of the rich
assortment of algebraic techniques for analysing matrices and linear systems.

The same mathematics can be applied to the local deformation of space produced
by the flow.  In this case finite matrices are an exact representation of the problem with the size equal to the dimension of the flow field.

When tracer transport is treated using matrix methods, there
are many special properties that can be derived for the solutions.  
Matrix methods are also
helpful in understanding and improving the stability of numerical solutions.

\section*{List of symbols}

\addcontentsline{toc}{section}{List of symbols}

\begin{tabular}{lll}
	Symbol & Description & First used \\ \hline
	$\vec x$ & spatial position & (\ref{trajectory_equation})\\
	$\vec v$ & velocity & (\ref{trajectory_equation})\\
	$t$ & time & (\ref{trajectory_equation})\\
	$\Phi$ & integrated trajectory & (\ref{traj_def}) \\
	$\Delta t$ & change in time; time step & (\ref{traj_def}) \\
	$\delta V$ & volume of integration & (\ref{volume_conservation_integral})\\
	$A$ & surface area of integration enclosing $\delta V$ & (\ref{volume_conservation_integral})\\
	$\rho$ & tracer density & (\ref{volume_conservation_integral})\\
	$\sigma$ & source term & (\ref{volume_conservation_integral})\\
	$q$ & tracer mixing ratio & (\ref{mixing_ratio})\\
	$\rho_t$ & fluid density & (\ref{mixing_ratio})\\
	$\delta \vec x$ & positional error & (\ref{taylor_expansion})\\
	$H$ & tangent model; deformation matrix & (\ref{deformation_matrix})\\
	$\vec x_0$ & initial position; Lagrangian coordinate & (\ref{initial_error_vector})\\
	$H^\prime$ & inverse deformation matrix & (\ref{inverse_deformation_matrix})\\
	$Q$ & continuous transport map & (\ref{tracer_map_continuous}) \\
	$V$ & volume of integration (whole space) & (\ref{tracer_map_continuous}) \\
	$R=\lbrace r_{ij} \rbrace$ & discrete transport map; solution matrix & (\ref{discrete_tracer_map0}) \\
	$A=\lbrace a_{ij} \rbrace$ & evolution matrix: coefficients in system of linear ODEs & (\ref{discrete_tracer_map}) \\
	$\Delta x$ & grid size in Eulerian tracer simulation & (\ref{simple_finite_difference}) \\
	$\Lambda = \lbrace \lambda_i \rbrace$ & diagonal matrix of eigenvalues & (\ref{eigenvalue_expansion}) \\
	$T$ & matrix of eigenvectors & (\ref{eigenvalue_expansion}) \\
	$\vec r$ & coordinates in a system of linear ODEs & (\ref{linear_ODE_system_vector_soln}) \\
	$U$ & matrix of left singular vectors & (\ref{SVD_def}) \\
	$S=\lbrace s_i \rbrace$ & diagonal matrix of singular values & (\ref{SVD_def}) \\
	$V=\lbrace v_{ij} \rbrace$ & matrix of right singular vectors & (\ref{SVD_def}) \\
	$m$ & number of rows in matrix & (\ref{SVD_def}) \\
	$n$ & number of columns; size/dimension of problem & (\ref{SVD_def}) \\
	$\lbrace h_i \rbrace$ & Lyapunov spectrum & (\ref{Lyap_def}) \\
	$D$ & diffusivity tensor & (\ref{advection_diffusion}) \\
	$d$ & diffusion coefficient & (\ref{finite_difference_diffusion}) \\
	$N$ & number of spatial dimensions & \S \ref{practical_considerations} \\
	$\omega$ & angular frequency & (\ref{diffusion_solution})
\end{tabular}

\newpage

\addcontentsline{toc}{section}{References}
\bibliography{tracer.bib}

\begin{thebibliography}{}

\bibitem[\protect\astroncite{Acheson}{1990}]{Acheson1990}
Acheson, D. (1990).
\newblock {\em Elementary Fluid Dynamics}.
\newblock Oxford University Press.

\bibitem[\protect\astroncite{Anderson}{1994}]{Anderson1994}
Anderson, Jr., J.~D. (1994).
\newblock {\em Computational Fluid Dynamics: The Basics with Applications}.
\newblock McGraw Hills.

\bibitem[\protect\astroncite{Anton}{1987}]{Anton1987}
Anton, H. (1987).
\newblock {\em Elementary Linear Algebra}.
\newblock Wiley.

\bibitem[\protect\astroncite{Cannon}{1984}]{Cannon1984}
Cannon, J.~R. (1984).
\newblock The {O}ne-{D}imensional {H}eat {E}quation.
\newblock volume~23 of {\em Encyclopedia of Mathematics and Its Applications}.
  Addison-Wesley.

\bibitem[\protect\astroncite{Courant et~al.}{1967}]{Courant_etal1967}
Courant, R., Friedrichs, K., and Lewy, H. (1967).
\newblock On the partial difference equations of mathematical physics.
\newblock {\em Journal of Research and Development}, 11(2):215--234.

\bibitem[\protect\astroncite{Katznelson}{1976}]{Katznelson1976}
Katznelson, Y. (1976).
\newblock {\em An Introduction to Harmonic Analysis}.
\newblock Dover.

\bibitem[\protect\astroncite{Mills}{2004}]{Mills2004}
Mills, P. (2004).
\newblock Following the vapour trail: a study of chaotic mixing of water vapour
  in the upper troposphere.
\newblock Master's thesis, University of Bremen.

\bibitem[\protect\astroncite{Mills}{2018}]{Mills2018}
Mills, P. (2018).
\newblock P{C} proxy: A method for dynamical tracer reconstruction.
\newblock {\em Journal of Environmental Fluid Mechanics}.
\newblock DOI:10.1007/s10652-018-9615-7.

\bibitem[\protect\astroncite{Ott}{1993}]{Ott1993}
Ott, E. (1993).
\newblock {\em Chaos in Dynamical Systems}.
\newblock Cambridge University Press.

\bibitem[\protect\astroncite{Ottino}{1989}]{Ottino1989}
Ottino, J.~M. (1989).
\newblock {\em The Kinematics of Mixing: Stretching, Chaos and Transport}.
\newblock Cambridge University Press.

\bibitem[\protect\astroncite{Pattanayak}{2001}]{Pattanayak2001}
Pattanayak, A.~K. (2001).
\newblock Characterizing the metastable balance between chaos and diffusion.
\newblock {\em Physica D}, 148:1--19.

\bibitem[\protect\astroncite{Pedlosky}{1987}]{Pedlosky1987}
Pedlosky, J. (1987).
\newblock {\em Geophysical Fluid Dynamics}.
\newblock Springer-Verlag, 2nd edition.

\bibitem[\protect\astroncite{Press et~al.}{1992}]{Press_etal1992}
Press, W.~H., Teukolsky, S.~A., Vetterling, W.~T., and Flannery, B.~P. (1992).
\newblock {\em Numerical Recipes in C}.
\newblock Cambridge University Press, 2nd edition.

\bibitem[\protect\astroncite{Robinson}{2004}]{Robinson2004}
Robinson, J.~C. (2004).
\newblock {\em An Introduction to Ordinary Differential Equations}.
\newblock Cambridge University Press.

\bibitem[\protect\astroncite{Thornton}{2003}]{Thornton2003}
Thornton, S.~T. (2003).
\newblock {\em Classical Dynamics of Particles and Systems}.
\newblock Brooks Cole, 5 edition.

\end{thebibliography}

\end{document}